\documentclass[aps,prd,twocolumn,superscriptaddress,nofootinbib,longbibliography,10pt]{revtex4-2}

\usepackage{graphicx}
\usepackage{amsfonts}
\usepackage{amsmath}

\usepackage[dvipsnames, usenames]{xcolor}
\definecolor{linkcolor}{rgb}{0.0,0.3,0.5}
\usepackage[unicode, colorlinks=true, linkcolor=linkcolor, citecolor=linkcolor, filecolor=linkcolor,urlcolor=linkcolor, pdfusetitle]{hyperref}

\begin{document}

\title{Bounds on the mass of superradiantly unstable scalar fields around Kerr black holes}

\author{Maur\'icio Richartz}
 \email{mauricio.richartz@ufabc.edu.br}
\affiliation{Centro de Matem\'atica, Computa\c c\~ao e Cogni\c c\~ao, Universidade Federal do ABC (UFABC), 09210-170 Santo Andr\'e, S\~ao Paulo, Brazil}

\author{Jo\~{a}o Lu\'{i}s Rosa}
\email{joaoluis92@gmail.com}
\affiliation{Institute of Theoretical Physics and Astrophysics, University of Gda\'nsk, Jana Ba\.{z}\'ynskiego 8, 80-309 Gda\'nsk, Poland}
\affiliation{Institute of Physics, University of Tartu, W. Ostwaldi 1, 50411 Tartu, Estonia}

\author{Emanuele Berti}
 \email{berti@jhu.edu}
\affiliation{Department of Physics and Astronomy, Johns Hopkins University, 3400 North Charles Street, Baltimore, Maryland 21218, USA}

\begin{abstract} 
In this work we compute numerical bounds on the mass $\mu$ of superradiantly unstable scalar fields in a Kerr black hole background using the continued fraction method. We show that the normalized upper bound on the mass $\mu$ increases with the angular momentum number $\ell$ and the azimuthal number $m$, approaching the most stringent analytical bound known to date when $\ell=m \gg 1$. We also provide an analytical fit to the numerically determined mass bound as a function of the dimensionless spin parameter $a/M$ of the black hole with an accuracy of the order $0.1\%$ for the fundamental mode with $\ell=m=1$, and of the order $1\%$ for higher-order modes (up to $\ell=m=20$). We argue that this analytical fit is particularly useful in astrophysical scenarios, since the lowest $\ell=m$ modes are capable of producing the strongest observable imprints of superradiance. 
 
\end{abstract}

\maketitle

\section{Introduction}\label{sec:intro}

Superradiance is a radiation enhancement process involving rotating dissipative systems~\cite{1971JETPL..14..180Z,1972JETP...35.1085Z}. It consists of the extraction of  rotational energy 
by the scattering of incoming low-frequency waves. As a result of superradiance, scattered radiation is amplified with respect to incident radiation~\cite{Bekenstein:1998nt,Richartz:2009mi,Brito:2015oca}. Both dissipation and angular momentum are crucial ingredients for superradiant scattering, which is predicted to occur around astrophysical black holes~\cite{1972BAPS...17..472M,Starobinsky:1973aij,1974JETP...38....1S}, analogue black holes~\cite{Basak:2002aw,Richartz:2014lda,Torres:2016iee,Cardoso:2022yin} and other compact objects \cite{Richartz:2013unq,Cardoso:2015zqa,Day:2019bbh,Cardoso:2017kgn}.

In black holes, dissipation is associated with the presence of the event horizon, which can be understood as a  one-directional membrane within which everything, including light, gets trapped. Rotation and angular momentum are ubiquitous in astrophysical black holes, that can be assumed to be electrically neutral~\cite{Gibbons:1975kk}. In the context of general relativity (GR), such astrophysical objects are described by the Kerr metric~\cite{Kerr:1963ud,Visser:2007fj,Bambi:2011mj,Teukolsky:2014vca,Berti:2019tcy}:
 \begin{align}
& ds^2 = - \left(1 - \frac{2Mr}{\rho ^2} \right) dt^2 - \frac{4 M r a \sin ^2 \theta}{\rho^2} d\phi \, dt  
+ \frac{\rho^2}{\Delta}dr^2 \nonumber \\ & \ \ \ + \rho^2 d \theta^2 + \left( r^2 + a^2 + \frac{2 M r a^2 \sin^2 \theta}{\rho ^2} \right) \sin ^2 \theta \, d \phi ^2, \label{metric}
\end{align}
where $M$ and $a \le M$ are the mass and the specific angular momentum of the black hole (here and throughout the paper we use natural units $G=c=\hbar=1$). The coordinates $x^a=(t,r,\theta,\phi)$ are named after Boyer and Lindquist. The auxiliary functions appearing in the metric are $\rho^2 = r^2 + a^2 \cos ^2 \theta$ and $\Delta = r^2 - 2Mr + a^2$. The roots of the equation $\Delta=0$, namely $r_+=M+\sqrt{M^2 - a^2}$ and $r_-=M-\sqrt{M^2 - a^2}$, correspond to the event horizon and to the Cauchy horizon of the black hole, respectively. The parameter $\Omega=a/2Mr_+$ is the angular velocity of the event horizon.  

As physically realistic objects, black holes must be stable against exterior perturbations, at least on cosmological timescales. Within GR, the stability of these objects has been studied for scalar, vector and tensor perturbations, and there is strong evidence that the subextremal Kerr black hole is stable against massless perturbations~\cite{Teukolsky:1972my,Whiting:1988vc,Dias:2015wqa} (see \cite{Aretakis:2011gz,Aretakis:2012bm,Lucietti:2012sf,Richartz:2015saa,Casals:2016mel,Richartz:2017qep} for analyses of the extremal case, and \cite{Rosa:2020uoi,Wondrak:2018fza,Alexander:2022avt} for analyses beyond GR). However, when the black hole is surrounded by certain confining media, black hole superradiance can give rise to exponentially growing modes which lead to superradiant instabilities. Such unstable configurations are known as ``black hole bombs''~\cite{Press:1972zz,Cardoso:2004nk}. 
The scattering of massive fields by rotating black holes is a notorious example of a black hole bomb in which the mass of the field acts as a mirror, effectively confining superradiant modes and leading to floating orbits~\cite{Cardoso:2011xi} or energy extraction from the black hole~\cite{Brito:2015oca}. Recently, superradiant instability calculations were used to derive upper bounds on the masses of the photon~\cite{Pani:2012vp} and of the graviton~\cite{Brito:2013wya}, to constrain dark matter models of dark photons and axion-like particles~\cite{Cardoso:2018tly}, and to investigate observational signatures of primordial magnetic black holes~\cite{Pereniguez:2024fkn}.

Superradiant instabilities of a massive scalar field around a Kerr black hole depend only on two quantities~\cite{Detweiler:1980uk,Dolan:2007mj,Dolan:2012yt}: the product $m\Omega$ between the azimuthal number $m$ of the field and the black hole angular velocity $\Omega$, and the product $\mu M$ between the scalar field mass $\mu$ and the black hole mass $M$. The former quantity is equal to the maximum frequency beyond which superradiance is hindered for a given $m$-mode, while the latter quantity corresponds to the ratio between the typical length scale associated with the black hole and the Compton wavelength of the scalar field.

Interestingly enough, the instability shuts down for sufficiently heavy fields, i.e., the scalar field is stable whenever its mass is larger than a certain critical mass $\mu_c$~\cite{Beyer:2000fz}. Analytical estimates of this upper mass bound for superradiant instabilities have improved over the years~\cite{Beyer:2011py,Hod:2012zza}. To date, the most stringent analytical bound on the mass of a stable scalar field, for a given black hole, is~\cite{Hod:2016iri}
\begin{equation}\label{bound}
\mu>m\Omega f(\gamma),
\end{equation}
where $\gamma=r_-/r_+$ and the function $f$ is defined by
\begin{equation} \label{fdef}
f(\gamma)=\gamma^{-\frac{3}{2}}\sqrt{2\left(1+\gamma\right)\left(1-\sqrt{1-\gamma^2}\right)-\gamma^2}.
\end{equation}

In this work, we compute numerically the mass bound $\mu_c$ for the stability threshold and compare the results with the analytical bound of Eq.~\eqref{bound}. We also provide a simple model to fit the numerical data.

The paper is organized as follows. In Sec.~\ref{sec:dynamics} we review the equations of motion for the perturbative scalar field in a Kerr background and make some considerations regarding its stability. In Sec.~\ref{sec:numerics} we compute numerically the upper bounds on the mass of the unstable scalar field using the continued fraction method, and we provide an analytical function to fit the associated data. In Sec.~\ref{sec:concl} we make some final remarks.

\section{Field dynamics}\label{sec:dynamics}

Scalar perturbations in a curved background are represented by a complex scalar field $\Phi$ which is minimally coupled to gravity through the Einstein-Hilbert action
\begin{equation}\label{action}
S= \int\sqrt{-g}\left[\frac{R}{8\pi}-\frac{1}{2}g^{ab}\partial_a\Phi^*\partial_b\Phi-\frac{\mu^2}{2}\Phi^*\Phi\right]d^4x,
\end{equation}
where $g$ is the determinant of the metric $g_{ab}$ and $R$ is the Ricci scalar. 
Varying the action with respect to the scalar field $\Phi$ yields the Klein-Gordon (KG) equation
\begin{equation}\label{kleingordon}
\left(\nabla^a\nabla_a-\mu^2\right)\Phi=0.
\end{equation}
Given the metric of a Kerr black hole, Eq.~\eqref{metric}, one can verify that the KG equation is separable for axisymmetric perturbations described by the ansatz~\cite{Brill:1972xj}
\begin{equation}\label{ansatz}
\Phi\left(t,r,\theta,\phi\right)=\frac{\psi\left(r\right)}{\sqrt{r^2 + a^2}}S\left(\theta\right)e^{-i\omega t+i m \phi},
\end{equation}
where the functions $\psi$ and $S$ depend solely on $r$ and $\theta$, respectively, and $\omega$ is the angular frequency of the scalar field. Note that the azimuthal number $m\in \mathbb Z$ is associated with an $m$-fold symmetry of the scalar field with respect to the polar angle $\phi$.

The separation of variables yields an angular equation for the function $S$ and a radial equation for the function $\psi$. The angular equation is a spheroidal wave equation~\cite{Berti:2005gp} 
\begin{equation}
\frac{1}{\sin \theta} \frac{d}{d \theta} \left( \sin \theta \frac{dS}{d \theta} \right) + \left(c^2 \cos^2 \theta + \lambda - \frac{m^2}{\sin ^2 \theta}   \right) S = 0, 
\end{equation}
where $c^2=a^2\left(\omega^2-\mu^2\right)$ and $\lambda$ is a separation constant. Regularity at the poles $\theta = 0$ and $\theta = \pi$ implies that only a discrete set of separation constants $\lambda=\lambda_{\ell m}(c)$ is allowed, where $\ell \ge |m|$ is a positive integer. The corresponding angular functions $S (\theta) = S_{\ell m}(\theta,c)$ are the spheroidal harmonics~\cite{Berti:2005gp}. In the regime of superradiant instabilities that we are interested in, the parameter $c$ is small, and $\lambda_{\ell m}(c) = \ell(\ell +1) + \mathcal O\left(c^2\right)$.

The radial equation, on the other hand, is given by
\begin{equation}\label{wave}
\frac{d^2\psi}{dr_*^2}+\left[\omega^2-V\left(\omega,r_*\right)\right]\psi=0,
\end{equation}
where $r_*$ is the tortoise coordinate defined by 
\begin{equation}
\frac{dr}{dr_*}=\frac{\Delta}{r^2+a^2}
\end{equation}
and the potential $V$, in terms of the original radial coordinate, is 
\begin{align}\label{potential}
V\left(\omega,r\right) = &\frac{2 a m \omega}{r^2 + a^2} +\frac{\Delta (\lambda + \mu^2 r^2) - a^2 m^2}{(r^2+a^2)^2} \nonumber \\
& + \frac{r \Delta \Delta'}{(r^2 + a^2)^3} - \frac{\Delta^2 (2r^2- a^2)}{(r^2 + a^2)^4}.
\end{align}

At the horizon $r=r_+$, for $\omega \in \mathbb{R}$, the solution of Eq.~\eqref{wave} is a linear combination of complex waves 
\begin{equation} \label{asymp1}
\psi\left(r\right)=A_+ e^{i\left(\omega-m\Omega\right)r_*}+B_+e^{-i\left(\omega-m\Omega\right)r_*},
\end{equation}
where $A_+$ and $B_+$ are the amplitudes of the outgoing and ingoing modes, respectively. Since the horizon is a one-directional membrane, the amplitude $A_+$ of the outgoing wave  must vanish. On the other hand, at infinity, the general solution has the form 
\begin{equation} \label{asymp2}
\psi\left(r\right)=A_\infty e^{i\sqrt{\omega^2-\mu^2}r_*} +B_\infty e^{-i\sqrt{\omega^2-\mu^2}r_*},
\end{equation}
where $A_\infty$ and $B_\infty$ are constants. If $\omega$ is real-valued and satisfies $\omega > \mu$, the solution far away from the black hole is also a combination of ingoing and outgoing modes. However, if the scalar field is sufficiently massive, we have exponentially growing and decaying modes. To avoid such growing modes, one sets the boundary condition $B_\infty = 0$.

For generic values of $\mu$, real-valued frequencies are not compatible with the boundary conditions introduced above. In fact, only a discrete set of complex frequencies $\omega=\omega_R+i\omega_I$ satisfy the boundary conditions $A_+=0$ and $B_\infty=0$ simultaneously for a given $\mu$. By solving the associated eigenvalue problem numerically, one typically finds stable solutions ($\omega_I<0$) that decay in time, known as quasibound states. Nevertheless, unstable modes ($\omega_I>0$) which grow in time are possible if $\mu$ is nonzero and sufficiently small. These modes correspond to superradiant instabilities since their real part satisfies the usual condition for superradiant scattering, i.e.,~$\omega_R < m\Omega$.

Bound states known as scalar clouds~\cite{Hod:2012px,Benone:2014ssa}, corresponding to confined modes with $\omega_I=0$, are also possible for a discrete set of masses $\mu_{sc}$. Such solutions oscillate with frequency $\omega=\omega_R=m\Omega$, and they are related to the possibility of hairy black hole solutions in GR~\cite{Herdeiro:2014goa,Berti:2019wnn}. For a given set of parameters $(\ell,m,a,M)$ characterizing the scalar field and the black hole, there is a discrete set of clouds, indexed by the non-negative integer $n$. The integer $n$ is associated with the number of nodes in the corresponding wave function: as $n$ increases, the scalar field becomes lighter. The mass of the heaviest possible cloud, corresponding to $n=0$, is the critical mass $\mu_c$ above which superradiant instabilites are absent.

\begin{figure}[htb!]
    \centering
    \includegraphics[width=8.6cm]{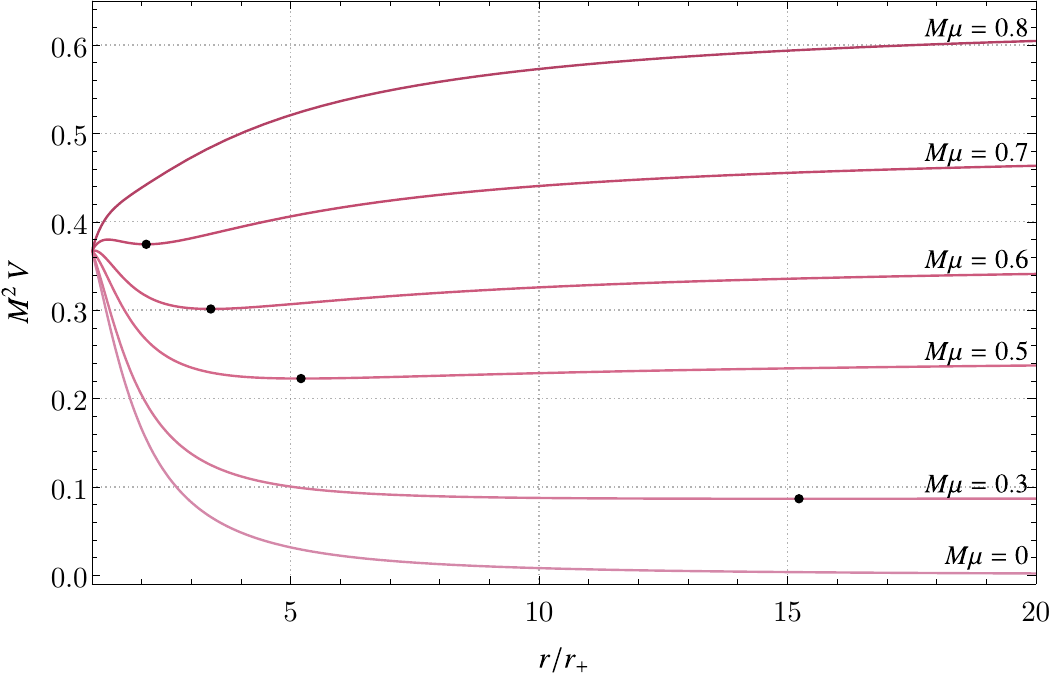}
    \caption{The effective potential $V(r)$, for $M\omega \in \mathbb{R}$, exhibiting the appearance of a potential well for certain values of the mass parameter $M\mu$. For illustration purposes, we have chosen the parameters that characterize the scalar field as $\ell=m=1$ and $M\omega = 1$, with $M\mu$ varying from zero (bottom curve) to $0.8$ (top curve). The spin of the black hole is set to $a/M=0.7$. Each black dot corresponds to the location of the minimum of the corresponding potential well. Note that for a massless field and for a sufficiently massive field (e.g., $M\mu  = 0.8$ in this example), the effective potential does not exhibit a potential well.}     
    \label{fig:potential}
\end{figure}

Intuitively, the fact that superradiant instabilities occur for massive fields can be identified with the fact that the potential $V(r)$, for fixed $\omega \in \mathbb{R}$, exhibits a minimum at some finite radius when $\mu$ is nonzero and sufficiently small, implying the existence of a potential well that confines the superradiant modes. The existence of a critical mass beyond which unstable modes are not allowed, on the other hand, can be graphically related to the behavior of the potential well as the mass of the scalar field increases from zero. Interestingly, despite an initial increase of the depth of the potential well, at some point the process comes to a halt and reverses direction. Eventually, when the mass $\mu$ is sufficiently large, the potential well disappears completely, eliminating the possibility of a confinement of superradiant modes~\cite{Hod:2012zza}. This behavior is illustrated in Fig.~\ref{fig:potential}.   

\section{Numerics and Results}\label{sec:numerics}

In this paper, we investigate the behavior of the critical mass $\mu_c$ as a function of the spin $a$ of the black hole for different perturbation modes. Based on the results of Ref.~\cite{Hod:2016iri}, which are summarized in Eq.~\eqref{bound}, we expect the critical mass $\mu_c$ of all $(\ell,m)$ modes to satisfy the bound
\begin{equation}\label{bound2}
\frac{\mu_c}{m \Omega} \le f(\gamma),
\end{equation}
where $f(\gamma)$, given in Eq.~\eqref{fdef}, is a function of $a/M$.

To determine $\mu_c$ we use the continued fraction method devised by Leaver in the context of black hole perturbation theory~\cite{Leaver:1985ax}. Since then, the method has been applied to a variety of black hole spacetimes~\cite{Berti:2009kk,Konoplya:2011qq}, and it is generally considered to be the gold standard technique to obtain the quasinormal modes and quasibound states of black holes. For the details on the implementation of the method, especially in the case of massive scalar fields around Kerr black holes, we refer the reader to Ref.~\cite{Dolan:2007mj}. 

Briefly, to compute superradiant instabilities (and also quasinormal modes and quasibound states) for a given set of parameters $(\ell,m,a,M)$, one fixes the mass $\mu$ and solves the associated continued fraction equation for the eigenvalue $\omega$. On the other hand, to compute the masses of the scalar clouds for a given set of parameters $(\ell,m,a,M)$, one fixes the frequency $\omega = m \Omega$ and solves the associated continued fraction equation  for the eigenvalue $\mu=\mu_{sc}$. Let $\mu_{sc}^{N}$ denote the solution of the continued fraction equation with $N$ terms. The number of terms $N$ used in our calculations is chosen so that   
\begin{equation}
\left| \frac{\mu_{sc}^{N+100} - \mu_{sc}^{N}}{\mu_{sc}^N}   \right| \le 10^{-10}.
\end{equation}

The algorithm used to compute the critical value $\mu_c$ as a function of the parameter $a/M$ for a given pair $(\ell,m)$ is divided in two main parts. The first part consists of determining, for $a/M=0.99999$, the heaviest scalar clouds. We do this by solving the continued fraction equation using several initial guesses for $M\mu$ and then sorting the solutions. The second part consists in following the first four resonances ($n=0,1,2,3$) as we successively decrease the spin parameter of the black hole. At each step (i.e.,~for a given spin parameter $a=a_0$), we interpolate the cloud masses that have already been found for $a>a_0$ to predict the cloud mass for $a=a_0$. Such predictions are used as initial guesses for the continued fraction method. At the end of each step, we store the values of $(\ell,m,n,a/M,M\mu_{sc})$. The algorithm terminates when $a/M=0.10000$.

\begin{figure}[t]
    \centering
    \includegraphics[width=8.6cm]{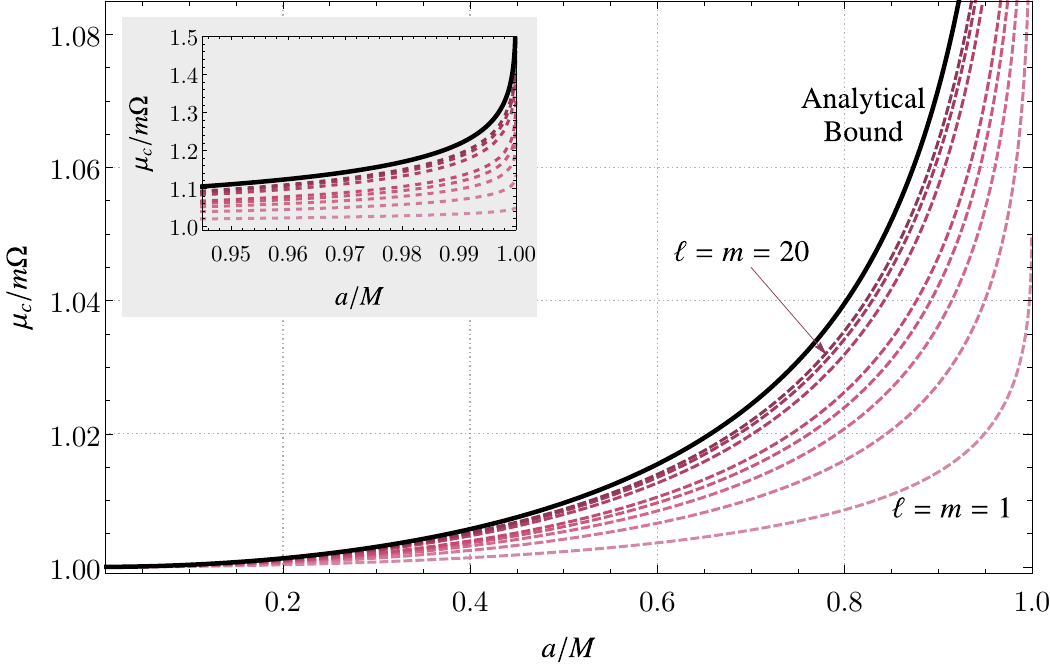}
    \caption{Normalized critical mass of the scalar field, $\mu_c/\left(m\Omega\right)$, as a function of the black hole spin $a/M$ for $\ell=m$ modes. From bottom to top, the dashed curves correspond to $\ell=m=1,2,3,4,5,10,15,20$. The black solid line is the function $f(\gamma)$ associated with the analytical bound given by Eq.~\eqref{bound}. Note that the critical values increase as $\ell=m$ increases. As $\ell=m\to \infty$, we expect the numerical results to asymptotically approach the analytical bound. The inset shows the same quantity over a broader range of $\mu_c/\left(m\Omega\right)$.}
    \label{fig:modes}
\end{figure}

\begin{figure}[t]
    \centering
    \includegraphics[width=8.6cm]{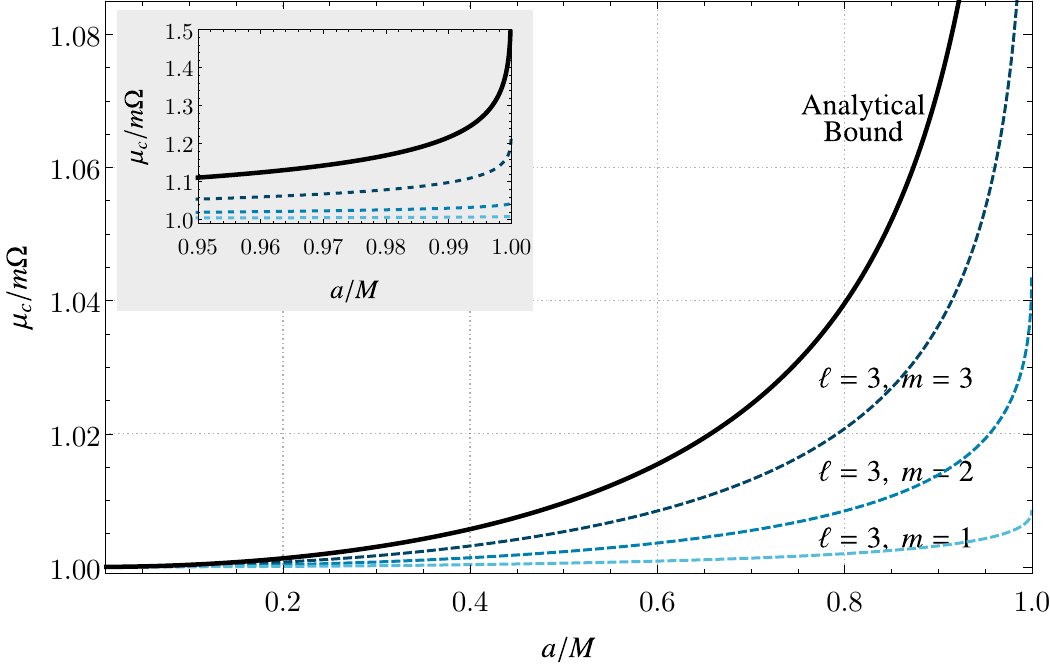}
    \caption{Normalized critical mass of the scalar field, $\mu_c/\left(m\Omega\right)$, as a function of the black hole spin $a/M$, for $\ell=3$ modes with different azimuthal number $m$ (dashed lines). The critical values increase as the azimuthal number $m$ increases for a fixed value of $\ell$. The solid black line is the function $f(\gamma)$ associated with the analytical bound given by Eq.~\eqref{bound}.}
    \label{fig:modes2}
\end{figure}

We can now apply the algorithm to several combinations of $\ell$ and $m$. The results for the critical mass $\mu_c$ (i.e.,~the mass of the heaviest scalar cloud) are shown in Figs.~\ref{fig:modes} and \ref{fig:modes2}. The associated data are available online at~\cite{repository}. In each figure, the dashed lines represent the critical value $\mu_c$ (normalized by the superradiant threshold $m\Omega$) as a function of the spin parameter $a/M$ of the black hole. In Fig.~\ref{fig:modes} we exhibit the $\ell=m$ modes to illustrate the effect of increasing $\ell$ and $m$ simultaneously. In Fig.~\ref{fig:modes2} we exhibit modes with fixed $\ell=3$ to illustrate the effect of increasing $m$ alone. The black solid line in both figures is the analytical bound given by Eq.~\eqref{bound}, which is independent of $\ell$ and $m$. As $\ell$ and $m$ increase, the curves representing the normalized critical mass move up, suggesting that the analytical bound is the asymptotic limit of the numerical results for $\ell \rightarrow \infty$ and $m \rightarrow \infty$. 

\begin{table}
\setlength{\tabcolsep}{8pt}
\renewcommand{\arraystretch}{1.5}
\begin{tabular}{c|c c c c}
\hline\hline 
$\left( l, m\right)$ & $p_1$ & $p_2$ & $p_3$ & $p_4$ \\
\hline
$\left( 1, 1\right)$ & $4.22296$ & $3.47889$ & $4.34825$ & $3.85449$\\
$\left( 2, 2\right)$ & $5.26660$ & $6.00154$ & $5.49001$ & $6.78058$\\
$\left( 3, 3\right)$ & $5.69461$ & $7.27908$ & $5.97788$ & $8.31672$\\
$\left( 4, 4\right)$ & $5.87961$ & $7.89133$ & $6.20263$ & $9.09382$\\
$\left( 5, 5\right)$ & $5.96960$ & $8.20542$ & $6.32105$ & $9.52186$\\
$\left( 10, 10\right)$ & $6.10655$ & $8.69856$ & $6.52905$ & $10.2963$\\
$\left( 15, 15\right)$ & $6.14120$ & $8.82101$ & $6.59282$ & $10.5353$\\
$\left( 20, 20\right)$ & $6.15598$ & $8.87159$ & $6.62343$ & $10.6498$\\
\hline\hline
\end{tabular}
\caption{Values of the free parameters $p_i$ in Eq.~\eqref{fitexp} for different combinations of $\ell$ and $m$.}
\label{tab:parameters}
\end{table}

\begin{figure}[htb!]
    \centering
    \includegraphics[width=8.6cm]{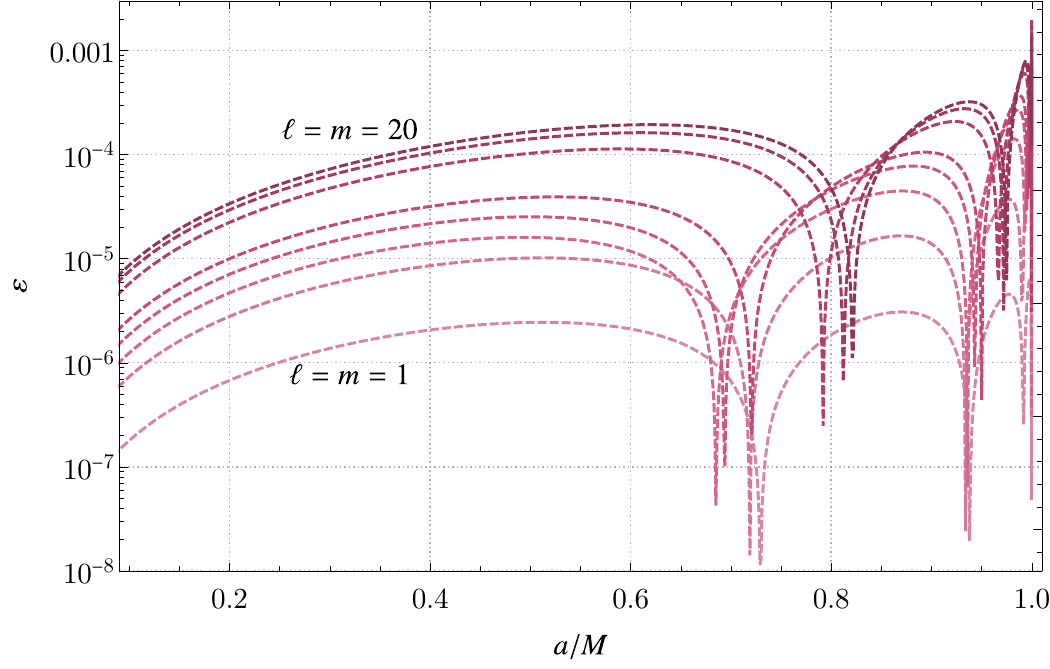}
    \caption{Absolute relative errors $\varepsilon$ of the fits in Eq.~\eqref{fitexp} with respect to the numerical values plotted in Fig.~\ref{fig:modes}. From bottom to top, the dashed curves correspond to $\ell=m=1,2,3,4,5,10,15,20$. The parameters used in Eq.~\eqref{fitexp} for each set of $\ell=m$ values are listed in Table~\ref{tab:parameters}.}
    \label{fig:errors}
\end{figure}

Even though there is no analytical exact formula for the mode-specific mass bounds shown in Fig.~\ref{fig:modes}, we were able to fit the data to relatively simple rational functions. We found that the relationship between the normalized critical value $\mu_c/\left(m\Omega\right)$ and the spin parameter $a$ can be approximated by a function of the form 
\begin{equation}\label{fitexp}
\frac{\mu_c^{\mathrm{fit}}}{m\Omega}=\frac{1-p_1(M \Omega)^2+p_2(M \Omega)^4}{1-p_3(M \Omega)^2+p_4(M \Omega)^4},
\end{equation}
where the $p_i$'s are free parameters which differ from case to case as we vary $\ell$ and $m$. In Table \ref{tab:parameters} we list the values of the parameters $p_i$ for different combinations of $\ell$ and $m$ (for other combinations of $\ell$ and $m$, see \cite{repository}).

In Fig.~\ref{fig:errors} we plot the absolute relative error $\varepsilon$ of the fits with respect to the numerical results.  The relative error is defined as
\begin{equation}
\varepsilon = \left| \frac{\mu_c - \mu_c^{\mathrm{fit}}}{\mu_c}   \right|,
\end{equation}    
where $\mu_c$ denotes the critical mass obtained through the continued fraction method, and $\mu_c^{\mathrm{fit}}$ denotes the critical mass determined by the analytical approximation in Eq.~\eqref{fitexp}. The relative error is always smaller than $0.001\%$ for the fundamental mode $\ell=m=1$, independently of the spin of the black hole. For higher-order modes, we have verified that the relative error is of order $0.1\%$ for rapidly spinning black holes ($a/M \sim 0.99$), whereas for slower black holes ($a/M \lesssim 0.5$) it is of order $0.01\%$.

The analytical bound in Eq.~\eqref{bound} and the fit in Eq.~\eqref{fitexp} should be useful in astrophysical scenarios. If superradiance operates in the Universe, most of its potentially observable consequences come from the $\ell=m=1$ mode, for large black hole spins. Since the inequality in Eq.~\eqref{bound} holds for all modes, we do not expect it to provide a good approximation for the lowest-$\ell$ modes. Indeed, as shown in Fig.~\ref{fig:modes}, the difference between the analytical bound and the critical mass for the lowest $\ell=m$ modes is large, especially for rapidly rotating black holes. In particular, at $a/M=0.99$, the relative difference is $17.8\%$ for the $\ell=m=1$ modes, and $13.7\%$ for the $\ell=m=2$ modes.   

\begin{figure}[htb!]
    \centering
    \includegraphics[width=8.6cm]{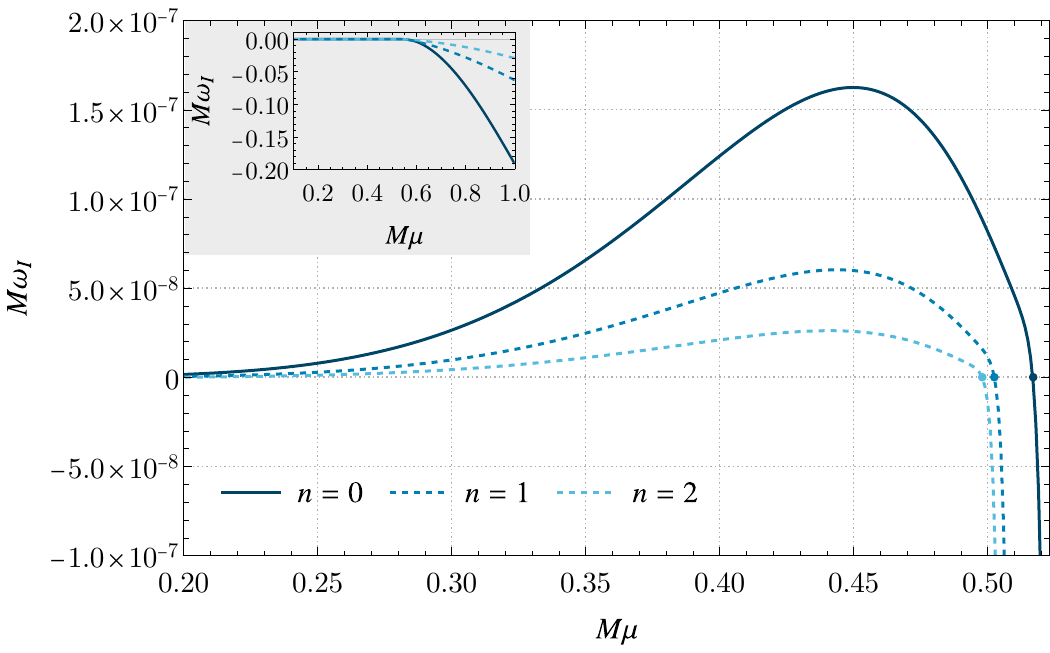}
    \caption{The imaginary parts $M\omega_I$ of the eigenvalues that satisfy the boundary conditions of quasibound states as a function of the scalar field's mass $\mu$ for three resonances of the mode $\ell=m=1$. The spin of the black hole is $a/M=0.9999$. The main plot focuses on the transition between stability and instability, which is indicated by the small circles and correspond to different scalar cloud configurations. The most unstable mode, represented by the solid curve, has the largest transition mass, which is the critical mass $\mu_c$. The resonances represented by the dashed curves, on the other hand, have the second and third largest transition masses. The inset is a zoomed-out plot of the transition, showing the behavior of the resonances over a larger range of the parameter $M\mu$.}
    \label{fig:example}
\end{figure}

\begin{figure}[htb!]
    \centering
    \includegraphics[width=8.6cm]{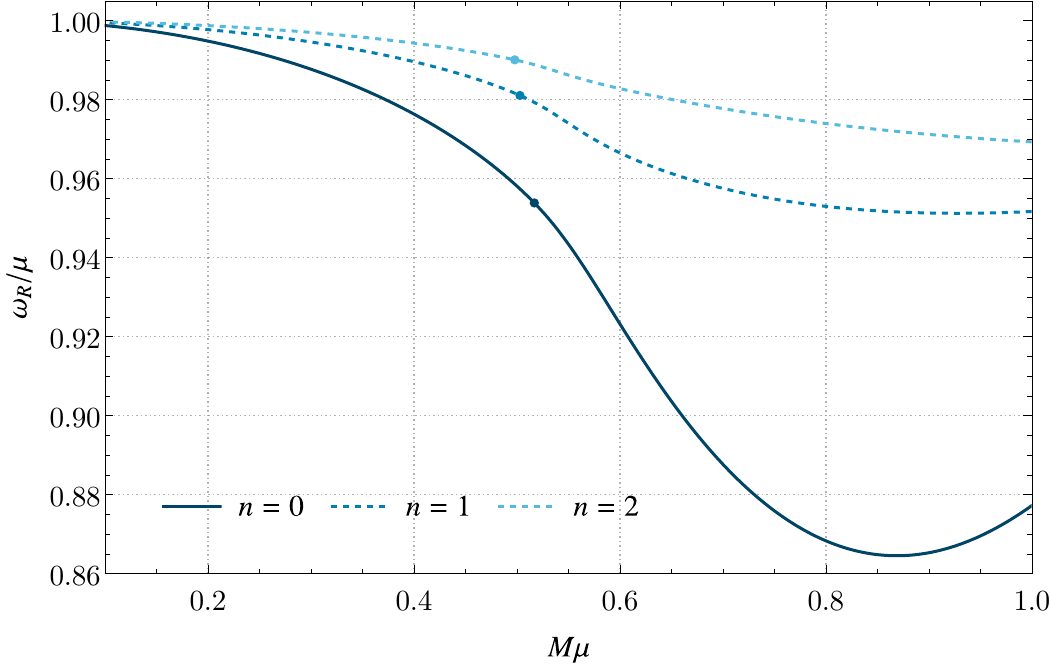}
    \caption{The real parts $\omega_R/\mu$ of the eigenvalues that satisfy the boundary conditions of quasibound states as a function of the scalar field mass $M\mu$ for three resonances of the mode $\ell=m=1$. The spin of the black hole is $a/M=0.9999$. The small circles correspond to different scalar cloud configurations, and indicate the real eigenvalues $\omega=\omega_R=m\Omega$ associated with the transition masses shown in Fig.~\ref{fig:example}.}
    \label{fig:example2}
\end{figure}

We close this section with a discussion of superradiant instabilities. Since scalar clouds correspond to transitions from quasibound states to superradiant instabilities, we can use the masses $\mu=\mu_{sc}$ and the corresponding frequencies $\omega= m \Omega$ as starting points to track superradiant instabilities as $\mu$ varies. In other words, for a given choice of $(\ell,m,n,a/M)$, we decrease $\mu$ incrementally from the starting value $\mu=\mu_{sc}$ and determine the corresponding eigenvalue $\omega$ in each step using the continued fraction method. As $\mu$ decreases from $\mu=\mu_{sc}$, the imaginary part of the frequency $\omega_I$ increases from zero to a maximum value $\omega_I{}_{\mathrm{max}}$. In Fig.~\ref{fig:example}, we exhibit such behavior for the first three resonances ($n=0,1,2$) of the mode $\ell=m=1$ when the spin is $a/M=0.9999$. The dots in Fig.~\ref{fig:example} mark the transition from stability to instability, indicating the scalar cloud masses above which the corresponding mode becomes stable. For completeness, in Fig.~\ref{fig:example2} we plot the real parts $\omega_R/\mu$ of the modes whose imaginary parts $M\omega_I$ are shown in Fig.~\ref{fig:example}.

\begin{figure*}[!htbp]
\centering
  \includegraphics[width = 0.99 \linewidth]{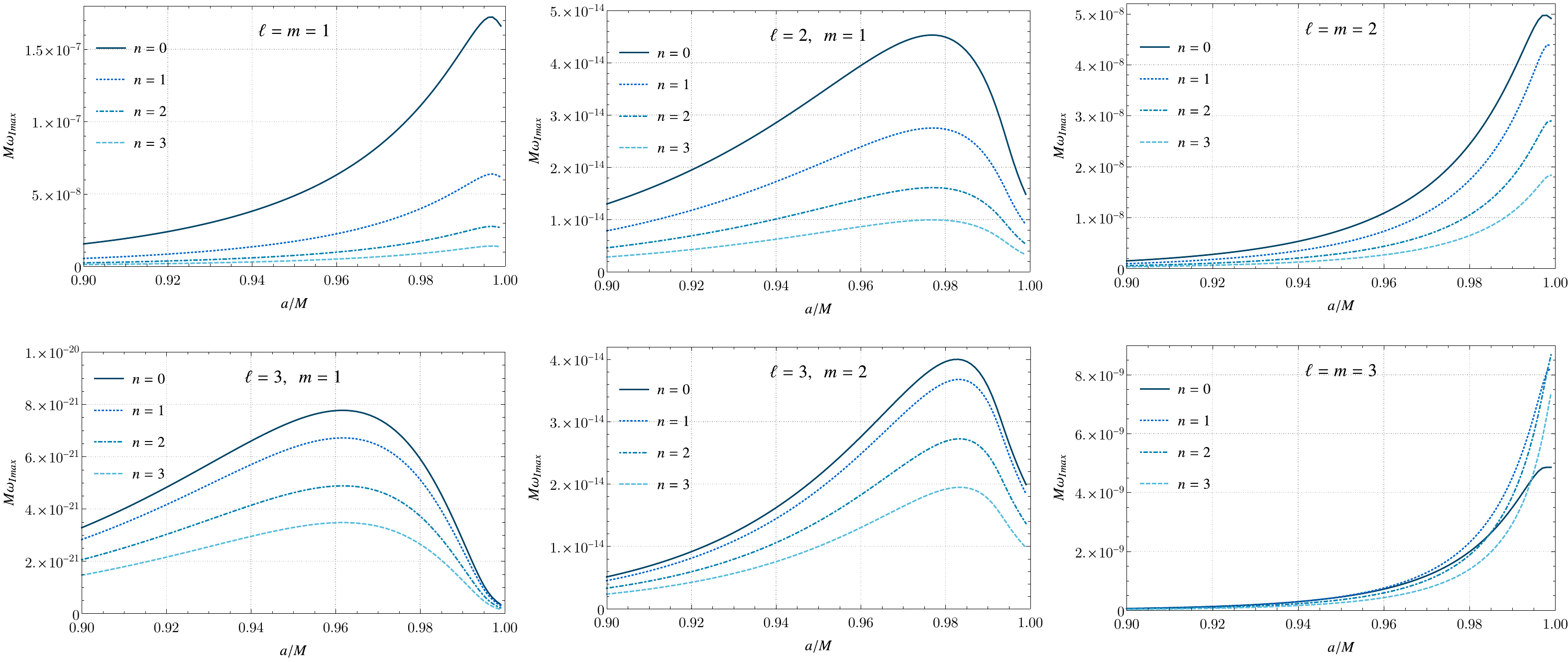}
  \caption{The maximum instability $M \omega_I{}_{\mathrm{max}}$ for $\ell \in \{1,2,3\}$ , $ 1 \le m \le \ell$ as a function of the spin $a/M$. The four curves in each panel are associated with the four heaviest scalar clouds (i.e.,~$n \in \{0,1,2,3\}$).
 }
  \label{fig:instabilitymaxima} 
\end{figure*}

The location of the maximum instability rate $\omega_I{}_{\mathrm{max}}$ as a function of the spin $a$ has been previously discussed in the literature for $\ell=m=1$~\cite{Dolan:2007mj,Dolan:2012yt,Ghosh:2018gaw}. In this work we provide the results for $\ell \in \{1,2,3\}$ when the spin lies in the range $0.2 \le a/M \le 0.999$. Since we start from the configuration corresponding to a scalar cloud and decrease the mass of the scalar field until the peak of the instability is found, each value $\omega_I{}_{\mathrm{max}}$ is associated a resonance number $n$.
The data for $\omega_I{}_{\mathrm{max}}$ (and for the corresponding values of oscillation frequency $\omega_R{}_{\mathrm{max}}$ and mass $\mu{}_{\mathrm{max}}$) are available online for $n \in \{0,1,2,3\}$~\cite{repository}. 
In Fig.~\ref{fig:instabilitymaxima} we show the values $M\omega_I{}_{\mathrm{max}}$ in the regime of high spins, where superradiance is stronger. We observe that, typically, as the mass of the scalar cloud increases (i.e., as $n$ decreases), $M\omega_I{}_{\mathrm{max}}$ increases. However, such behavior can be reversed for highly spinning black holes, as is the case for $\ell=m=3$ when $a/M \gtrsim 0.995$. 

\section{Final Remarks}\label{sec:concl}

In this work, we have used the continued fraction method to compute the critical values of the mass of a scalar perturbation field in the Kerr black hole background above which the configurations are superradiantly stable. We have also provided an analytical approximation of these values which approximates the numerical results with errors smaller than $0.1\%$. We have verified that the critical masses increase both with $\ell$ (for $\ell=m$) and with $m$ (at fixed $\ell$).

The strictest analytical bound proposed to date \cite{Hod:2016iri} is consistent with the numerical results, although it is a fairly conservative bound for modes with low values of $\ell$ and $m$. Furthermore, in the limit where $\ell$ and $m$ approach infinity, the numerically obtained critical masses asymptotically approach this analytical bound. In fact, the time scales associated with superradiant instabilites increase as $\ell=m$ increases. For instance, the $\ell=m=1$ mode is most unstable when the spin of the black hole is $a/M \approx 0.997$ and the mass of the scalar field is $M\mu \approx 0.449$. The corresponding instability timescale is $ (M \omega_I)^{-1}
\approx 5.8 \times 10^6$~\cite{Dolan:2007mj,Dolan:2012yt,Siqueira:2022tbc}. Therefore, for practical purposes, one might be interested in avoiding only the most unstable modes. Recognizing that potentially observable effects of superradiance would arise from the lowest $\ell=m$ modes, the analytical fit in Eq.~\eqref{fitexp} provides a useful approximation of the critical mass in astrophysical settings.   

As a future research direction, one could investigate analytical bounds for massive spin-1 and massive spin-2 fields around Kerr black holes to generalize the analytical bounds discussed in~\cite{Beyer:2000fz,Beyer:2011py,Hod:2012zza,Hod:2016iri} and the analytical fit proposed in this work. We remark that superradiant instabilities and stationary clouds associated with Proca fields have already been investigated in Refs.~\cite{Dolan:2018dqv,Santos:2020sut,Santos:2020pmh}. Similarly, superradiant instabilities for massive spin-2 fields were discussed in Refs.~\cite{Brito:2013wya,Brito:2020lup,Dias:2023ynv}.

Taking into account the Kerr hypothesis~\cite{Bambi:2011mj,Herdeiro:2022yle}, another interesting line of research would be to analyze how deviations from the Kerr metric affect the analytical bounds on the critical mass $\mu_c$. Previous work on superradiant instabilities and quasinormal modes of Kerr-like black holes could be a starting point for such an investigation~\cite{Glampedakis:2017dvb,Franzin:2021kvj,Siqueira:2022tbc}. In particular, for the nonrotating case, the presence of a cosmological constant seems to suppress the quasibound states~\cite{Correa:2024xki}, and it would be interesting to analyze how rotation affects this result.

\begin{acknowledgments}
J.L.R. acknowledges the European Regional Development Fund and the programme Mobilitas Pluss for financial support through Project No.~MOBJD647, project No.~2021/43/P/ST2/02141 co-funded by the Polish National Science Centre and the European Union Framework Programme for Research and Innovation Horizon 2020 under the Marie Sklodowska-Curie grant agreement No. 94533, Fundação para a Ciência e Tecnologia through project number PTDC/FIS-AST/7002/2020, and Ministerio de Ciencia, Innovación y Universidades (Spain), through grant No. PID2022-138607NB-I00. M.~R.~acknowledges partial support from the Conselho Nacional de Desenvolvimento Cient\'{i}fico e Tecnol\'{o}gico (CNPq, Brazil), Grant 315991/2023-2, and from the S\~ao Paulo Research Foundation (FAPESP, Brazil), Grant 2022/08335-0. 
E.B. is supported by NSF Grants No. AST-2006538, PHY-2207502, PHY-090003 and PHY-20043, by NASA Grants No. 20-LPS20-0011 and 21-ATP21-0010, by the John Templeton Foundation Grant 62840, by the Simons Foundation, and by the Italian Ministry of Foreign Affairs and International Cooperation grant No.~PGR01167. Part of this work was carried out at the Advanced Research Computing at Hopkins (ARCH) core facility (\url{rockfish.jhu.edu}), which is supported by the NSF Grant No.~OAC-1920103.
\end{acknowledgments}

\bibliography{boundsref}

\end{document}